\documentclass[prl,twocolumn,superscriptaddress]{revtex4}

\usepackage{graphicx}

\begin{document}

\title{Demonstrating a Driven Reset Protocol for a Superconducting Qubit}
\author{K. Geerlings}
\affiliation{Department of Applied Physics, Yale University, New Haven, Connecticut 06520-8284, USA}
\author{Z. Leghtas}
\affiliation{INRIA Paris-Rocquencourt, Domaine de Voluceau, B.P. 105, 78153 Le Chesnay cedex, France}
\author{I.M. Pop}
\author{S. Shankar}
\author{L. Frunzio}
\author{R.J. Schoelkopf}
\affiliation{Department of Applied Physics, Yale University, New Haven, Connecticut 06520-8284, USA}
\author{M. Mirrahimi}
\affiliation{Department of Applied Physics, Yale University, New Haven, Connecticut 06520-8284, USA}
\affiliation{INRIA Paris-Rocquencourt, Domaine de Voluceau, B.P. 105, 78153 Le Chesnay cedex, France}
\author{M.H. Devoret}
\affiliation{Department of Applied Physics, Yale University, New Haven, Connecticut 06520-8284, USA}

\date{October 19, 2012}

\begin{abstract}
Qubit reset is crucial at the start of and during quantum information algorithms. We present the experimental demonstration of a practical method to force qubits into their ground state, based on driving appropriate qubit and cavity transitions. Our protocol, called the double drive reset of population, is tested on a superconducting transmon qubit in a three-dimensional cavity. Using a new method for measuring population, we show that we can prepare the ground state with a fidelity of at least 99.5~$\%$ in less than 3~$\mu$s; faster times and higher fidelity are predicted upon parameter optimization.
\end{abstract}

\maketitle

	A method for qubit initialization is one of the fundamental requirements of quantum information processing laid out by DiVincenzo \cite{Divincenzo2000}.  Due to recent advancements in extending superconducting qubit relaxation times to the 100~$\mu$s range \cite{Paik2011a}, active ground state preparation (qubit reset), other than by passively waiting for equilibration with a cold bath, is becoming a necessity.  The main use for a fast, high-fidelity reset is to place the qubit into a known pure state either before or during an algorithm.  Active reset is preferred over passive reset when (a) the qubit thermal environment is hot on the scale of the transition frequency, and (b) rapid evacuation of entropy from the system is necessary, as in implementations of quantum error correction \cite{Schindler2011,Reed2012}.

	The ancestor of active qubit reset is dynamical cooling of nuclear spins using paramagnetic impurities  \cite{Abragam1961}.  Superconducting qubits are analogous to single spins in a controlled environment, and it is therefore possible to design similar dynamical cooling methods to achieve reset times much faster than the relaxation time $T_1$.  While several methods \cite{Valenzuela2006,Grajcar2008,Reed2010,Mariantoni2011,Riste2012,Johnson2012,Riste2012b,Campagne2012} for reset and dynamical cooling have been demonstrated in superconducting qubits, they each require either qubit tunability or some form of feedback and high-fidelity readout.  We present a practical dynamical cooling protocol without these requirements.  This protocol is related to dissipation engineering \cite{Poyatos1996}, as we use the dissipation through the cavity to stabilize the qubit ground state.  Double Drive Reset of Population (DDROP) is tested on a transmon qubit \cite{Schreier2008} in a three-dimensional cavity \cite{Paik2011a} but can be applied to any circuit QED system.

	DDROP consists of a pulse sequence that manipulates the transition landscape of the qubit-cavity system in order to quickly drive the qubit to the ground state.  The protocol relies on the number splitting property of the strong dispersive regime \cite{Schuster2007a} of circuit QED, where the dispersive shift $\chi$ of the cavity due to a qubit excitation is larger than twice the cavity linewidth $\kappa$ and qubit linewidth $1/T_{2}$.  Thus the cavity frequency depends on the state of excitation of the qubit, and the qubit frequency depends on the number of excitations in the cavity.  Another requirement is needed:  $\kappa$ must be much larger than $\Gamma_{up} = P_e/T_{1}$, where $P_e$ is the equilibrium excited state population.  This condition is easy to satisfy with the recent advances in extending $T_1$.  Apart from special cases where it is desirable to have small $\kappa$, most transmons and other qubits read by a superconducting cavity are candidates for this type of reset.

\begin{figure}
\begin{centering}
\includegraphics[width=3.375in]{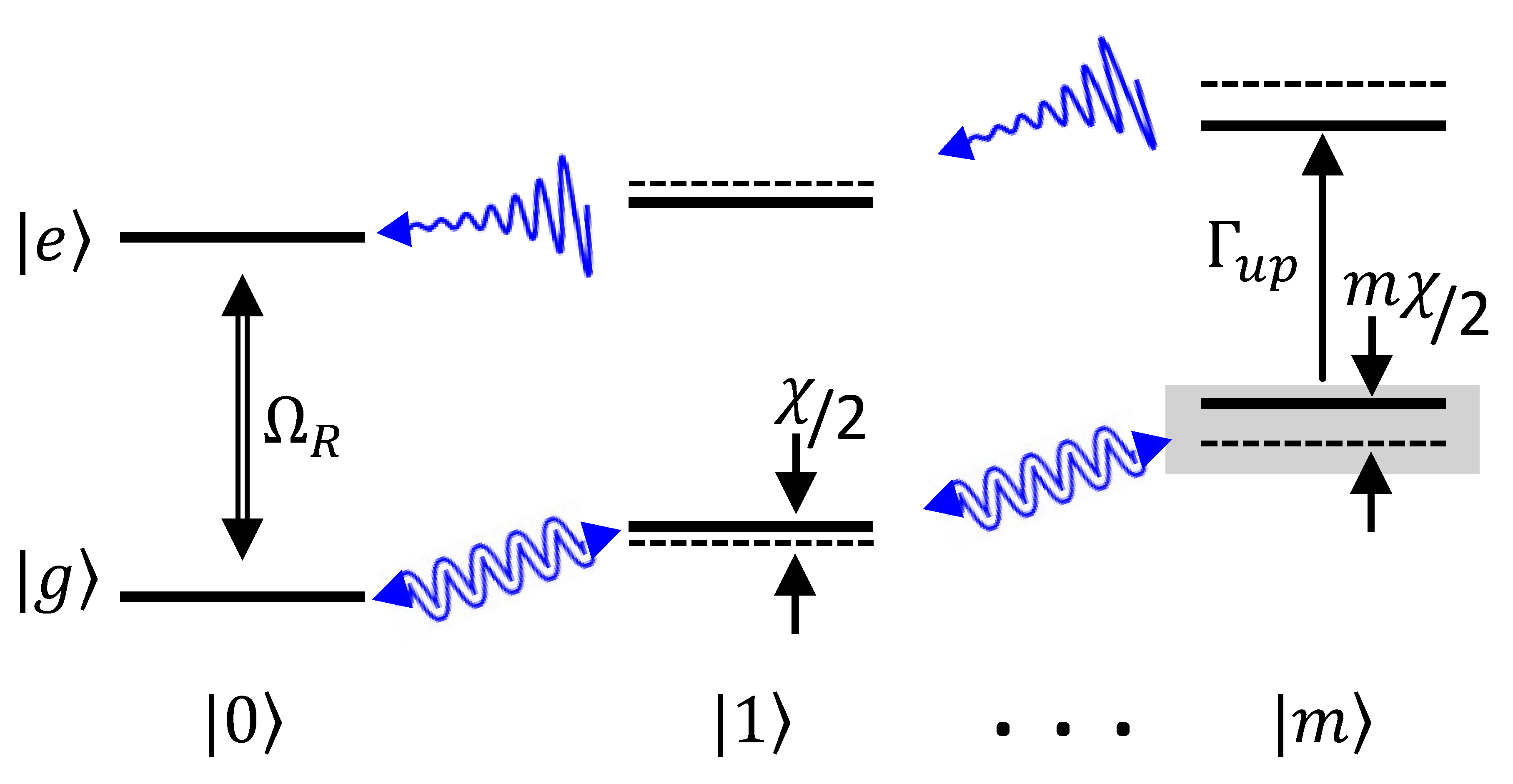}
\caption{Level structure of the transmon qubit coupled dispersively to a single resonator mode. The qubit excitations are spanned vertically while the resonator photon numbers are spanned horizontally.  The arrows show the transitions involved in the DDROP procedure along with their rates, with $\Gamma_{up} \ll \kappa \approx \Omega_{R} < \chi/2$.  The double arrows are driven transitions, while single arrows are spontaneous.  Qubit transitions are represented by straight lines while cavity transitions are wavy lines.  The steady-state equilibrium qubit-cavity joint state is the coherent state $\mid$$g,\alpha\rangle$.  For visualization, the state $\mid$$g,m\rangle$ is highlighted, where $m$ is the closest integer to the steady-state average number of photons in the cavity.}
\label{fig:reset_procedure}
\end{centering}
\end{figure}

	 In the DDROP protocol, shown graphically in Fig. \ref{fig:reset_procedure}, two microwave drives are applied simultaneously for a duration of order 10~$\kappa^{-1}$ in order to reach a steady state.  The first drive frequency, $f_{ge}^0$, is chosen in order to Rabi drive the qubit if the cavity has zero photons.  The amplitude of this drive is quantified by $\Omega_R$, the Rabi frequency.  The second frequency, $f_c^g$, is chosen to populate the cavity with photons if and only if the qubit is in the ground state.  The role of the cavity drive is to lift the population of $\mid$$g,0\rangle$ to the coherent state $\mid$$g,\alpha\rangle$, where $|\alpha|^2 = \bar{n}$, the steady state average photon number in the cavity.  Due to number splitting, the qubit transition frequencies when the cavity is in state $\mid$$\alpha\rangle$ differ sufficiently from $f_{ge}^0$ that the Rabi drive does not excite $\mid$$g,\alpha\rangle$.   The only way for the system to leave $\mid$$g,\alpha\rangle$ is through a spontaneous excitation happening at a rate $\Gamma_{up}$, which is slow compared to all other rates in the system.  Once in $\mid$$e,\alpha\rangle$, the system rapidly falls back to $\mid$$e,0\rangle$ in a time of order $\kappa^{-1}$.  The role of the Rabi drive, with Rabi frequency of order $\kappa$, is to speed up the transition between $\mid$$e,0\rangle$ and $\mid$$g,0\rangle$, thus allowing a fast return to $\mid$$g,\alpha\rangle$.  With both drives on, the system will be driven to  $\mid$$g,\alpha\rangle$ at a rate of order $\kappa$ regardless of initial state, while the rate $\Gamma_{up}$ away from this state is slow.  Eventually, to prepare $\mid$$g,0\rangle$ instead of $\mid$$g,\alpha\rangle$, one must turn off the drives and wait for the photons to decay in a time of several $\kappa^{-1}$.  Since the cavity is in a coherent state, this waiting time could be avoided by using a displacement pulse, which is easier to calibrate with cavities with higher quality factor.  The ratio $\kappa$/$\Gamma_{up}$ determines the fidelity of the ground state preparation, and must therefore be much greater than 1.

	The measurements presented here were performed in a standard circuit QED setup on an aluminum transmon qubit, fabricated using a bridgeless double-angle evaporation technique \cite{Rigetti2009,Lecocq2011}, inside a three-dimensional copper cavity, thermally anchored to the mixing chamber of a dilution refrigerator with a base temperature of 17~mK. The cavity was mounted inside a copper shield coated with infrared-absorbing material on the inside.  A high-frequency filter similar to that of Ref. \cite{Santavicca2008} and a microwave 12~GHz low-pass filter were placed on each input and output microwave line.  Two 8-12~GHz circulators were installed between the cavity and the HEMT amplifier.  System parameters were measured to be: $f_c^g$ = 9.1	~GHz, $f_{ge}^0$ = 5.0~GHz, $\kappa$/2 $\pi$ = 3~MHz, $\chi$/2 $\pi$ = 7~MHz, $T_1$ = 37~$\mu$s, $T_{2}^{Ramsey}$ = 20~$\mu$s, $T_{2}^{Echo}$ = 40~$\mu$s, equilibrium $P_e$ = 9~$\%$, $\Gamma_{up}/2\pi \approx$ 400~Hz.  Both requirements for the reset mechanism are achieved, with $\chi$/$\kappa$ = 2.3 and $\kappa$/$\Gamma_{up} \simeq$ 8,000.

\begin{figure}
\includegraphics[width=3.375in]{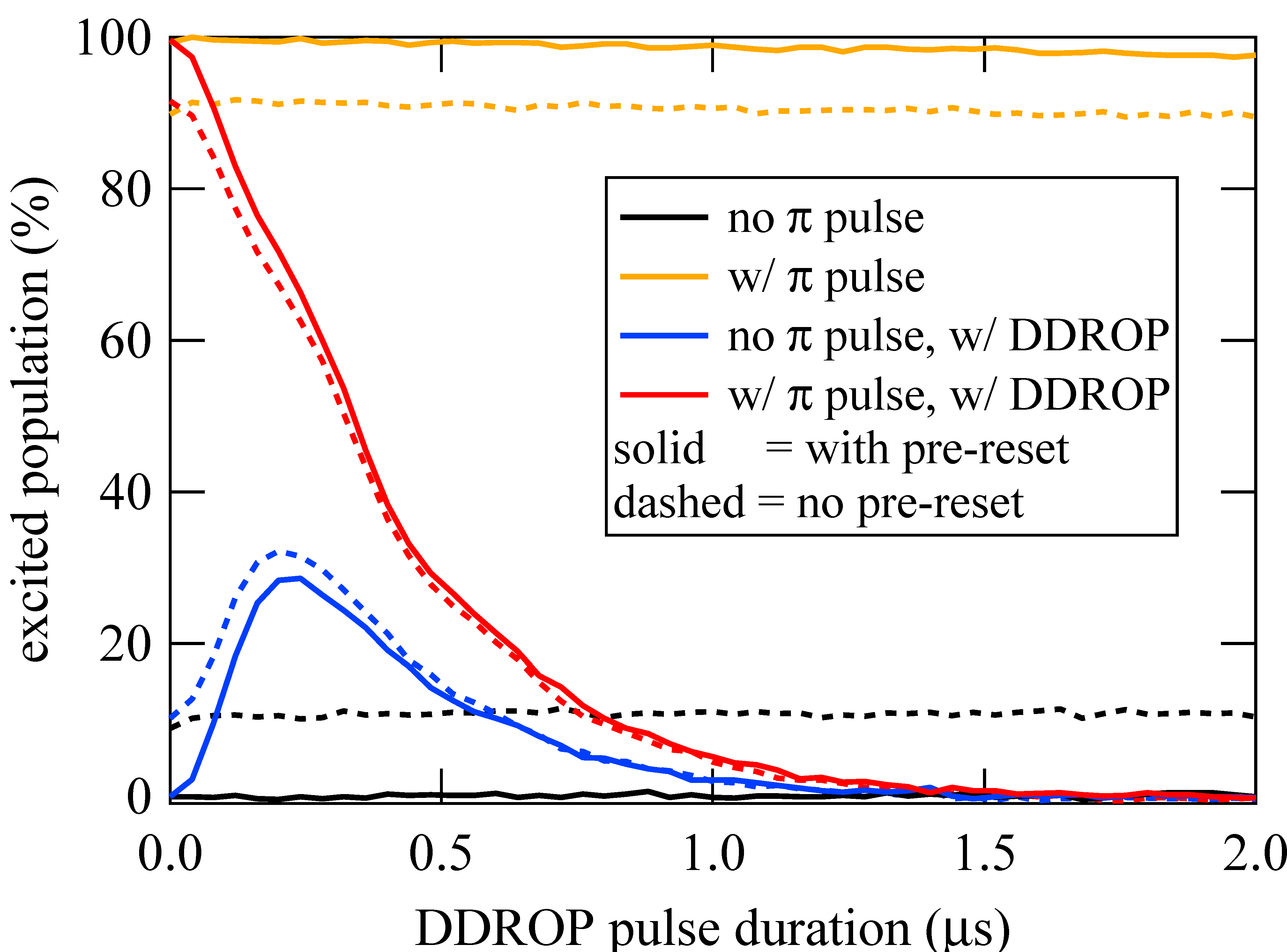}
\caption{Measured excited state population after reset pulse of varying duration, for four different initial preparations, measured after intervals of 40~ns.  The solid lines include a pre-reset while the dashed lines begin with the steady state 9$\%$ excited population. The `w/$\pi$ pulse' curve shows a slight downward trend due to the finite $T_1$.  The curves with DDROP show that, regardless of initial state, the qubit is driven to the ground state for pulse durations less than 2~$\mu$s.  For this measurement, $\Omega_R \approx$ 0.8~$\kappa$ and $\bar{n}$ = 8}
\label{fig:reset_measurement}
\end{figure}

	The effect of the DDROP protocol on this qubit is shown in Fig. \ref{fig:reset_measurement}, where the $y$~axis is the measured excited state population and the $x$~axis is the duration of the reset pulses (or delay time).  Each data point is taken after waiting 1~$\mu$s (20~$\kappa^{-1}$) after the end of the DDROP pulse to allow the system adequate time to decay from $\mid$$g,\alpha\rangle$ to $\mid$$g,0\rangle$.  The two solid, nearly horizontal curves are the pre-reset ground and excited qubit states without DDROP pulse.  The pre-reset is itself a 5~$\mu$s DDROP sequence done before all other pulses in order to suppress the initial excited state population.  The slight downward trend in the excited state curve, due to the finite value of $T_{1}$, is barely noticeable on this scale.  The other two solid curves correspond to the same preparation, but show the effect of a DDROP pulse whose duration is varied across the $x$~axis.  At short pulse duration, both initial populations tend towards 50$\%$ excitation, due to the Rabi drive.  As the duration is increased, the population tends quickly towards the pre-reset ground state.  The four dashed curves represent an identical set of data taken without the pre-reset, thus showing the effect of initial equilibrium population.  Note that regardless of the initial state, DDROP forces the population to the ground state in less than 3~$\mu$s (including the 1~$\mu$s decay from $\mid$$g,\alpha\rangle$ to $\mid$$g,0\rangle$).  This is a factor of 60 improvement over the standard protocol of waiting 5~$T_1$, which would give a comparable reduction of excited state population in a cold qubit environment.

	In order to benchmark our DDROP reset procedure, we had to carefully measure the resulting ensemble-averaged excited state population.  A measurement of the ratio of the heights of the two spectroscopic peaks corresponding to the $\mid$$g\rangle$ to $\mid$$e\rangle$ and $\mid$$e\rangle$ to $\mid$$f\rangle$ qubit transitions, usually assesses the excited state population.  However, this method does not take into account the variation of readout efficiency with qubit state, and is therefore not quantitative without further corrections.  

\begin{figure}
\includegraphics[width=3.375in]{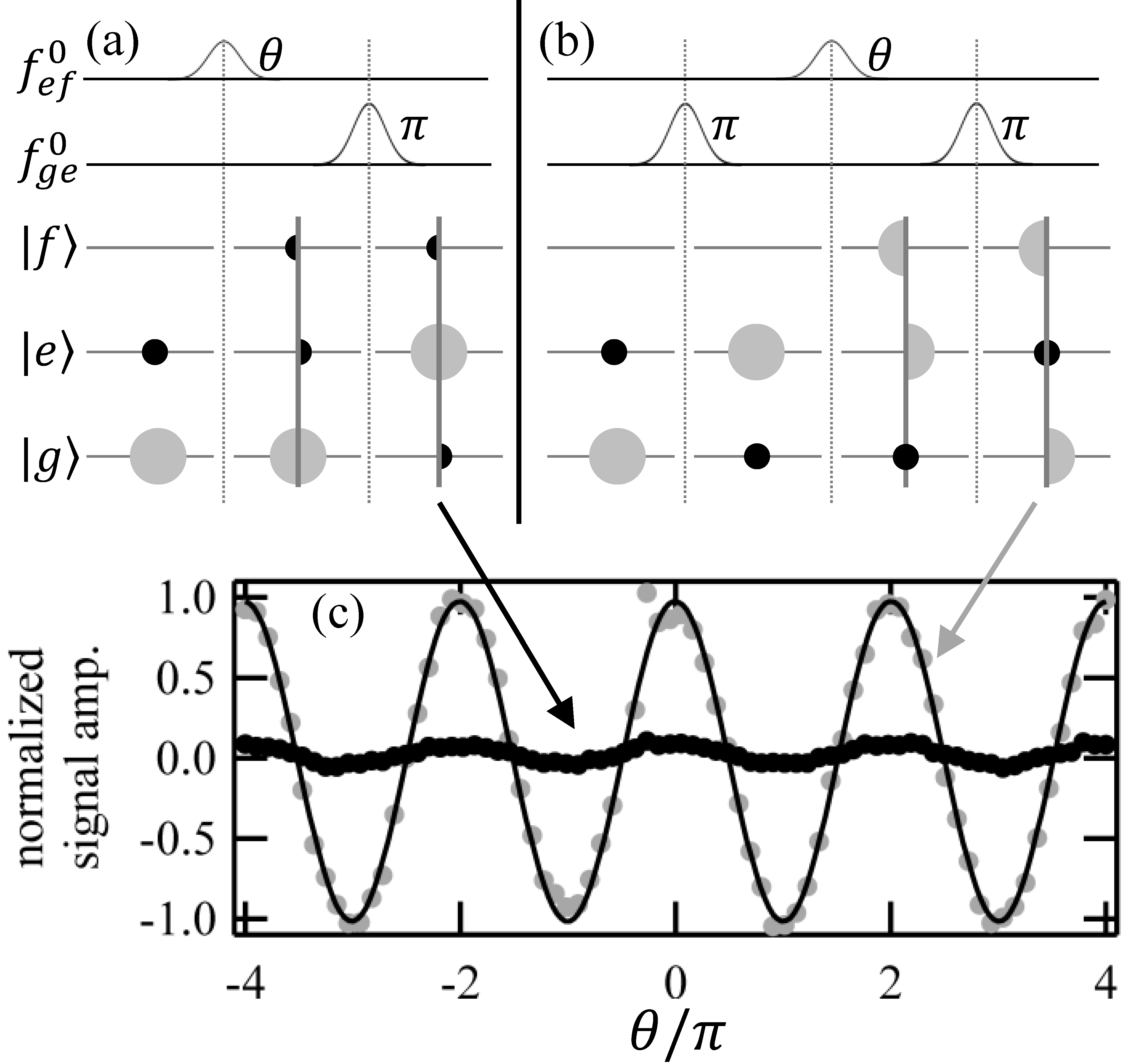}
\caption{Upper panel: pulse sequences used to perform qubit population measurement (RPM, see text), each producing an oscillation whose amplitude is proportional to initial excited (a) and ground (b) state population.  Circle radii indicate population in each state, vertical bars separate the two extrema in Rabi oscillations. Lower panel (c): example normalized data for measurement of 7$\%$ excited state population.}
\label{fig:population_measurement}
\end{figure}

	We introduce a method called the Rabi population measurement (RPM) that circumvents these problems.  The basic idea of RPM is to measure two Rabi oscillations whose amplitude ratio corresponds directly to the ratio of initial excited state ($P_e$) to ground state population ($P_g$).  This method is similar to, but different from, techniques previously used in phase qubits \cite{Lucero2008,Neeley2009}.  Note that for the cases treated in this paper, populations of states above $\mid$$e\rangle$ are negligible, so $P_e$ + $P_g$ = 1.  The RPM is performed by applying two sequences of qubit pulses as shown in Fig. \ref{fig:population_measurement}.  The first sequence consists of a pulse performing a rotation around $X$ on the $\mid$$e\rangle$ to $\mid$$f\rangle$ transition with varying angle $\theta \in [0,2\pi]$, followed by a $\pi$ pulse on the $\mid$$g\rangle$ to $\mid$$e\rangle$ transition.  Measuring the population of the $\mid$$g\rangle$ state results in a Rabi oscillation $A_e$cos($\theta$) with an amplitude $A_e$ proportional to $P_e$.  The second sequence differs only by the insertion of a $\pi$ pulse to first invert the population of the $\mid$$g\rangle$ and $\mid$$e\rangle$ states, yielding a Rabi oscillation $A_g$cos($\theta$) with an amplitude $A_g$ proportional to $P_g$.  The proportionality constants between the Rabi oscillation amplitudes and the corresponding populations are equal since the same transition is used in both sequences, thus avoiding readout efficiency variations.  From the two oscillation amplitudes, an estimate of the population and its associated standard deviation can be calculated from $P_e = A_e/(A_e+A_g)$.  The RPM protocol is self-calibrating and accesses smaller amplitudes than crude population measurements since it relies on the amplitude of an oscillation instead of just one value, in a lock-in fashion.  The minimum measurable value of $P_e$ was approximately 0.5$\%$, limited for technical reasons by the characteristics of our readout amplification chain.

\begin{figure}
\includegraphics[width=3.375in]{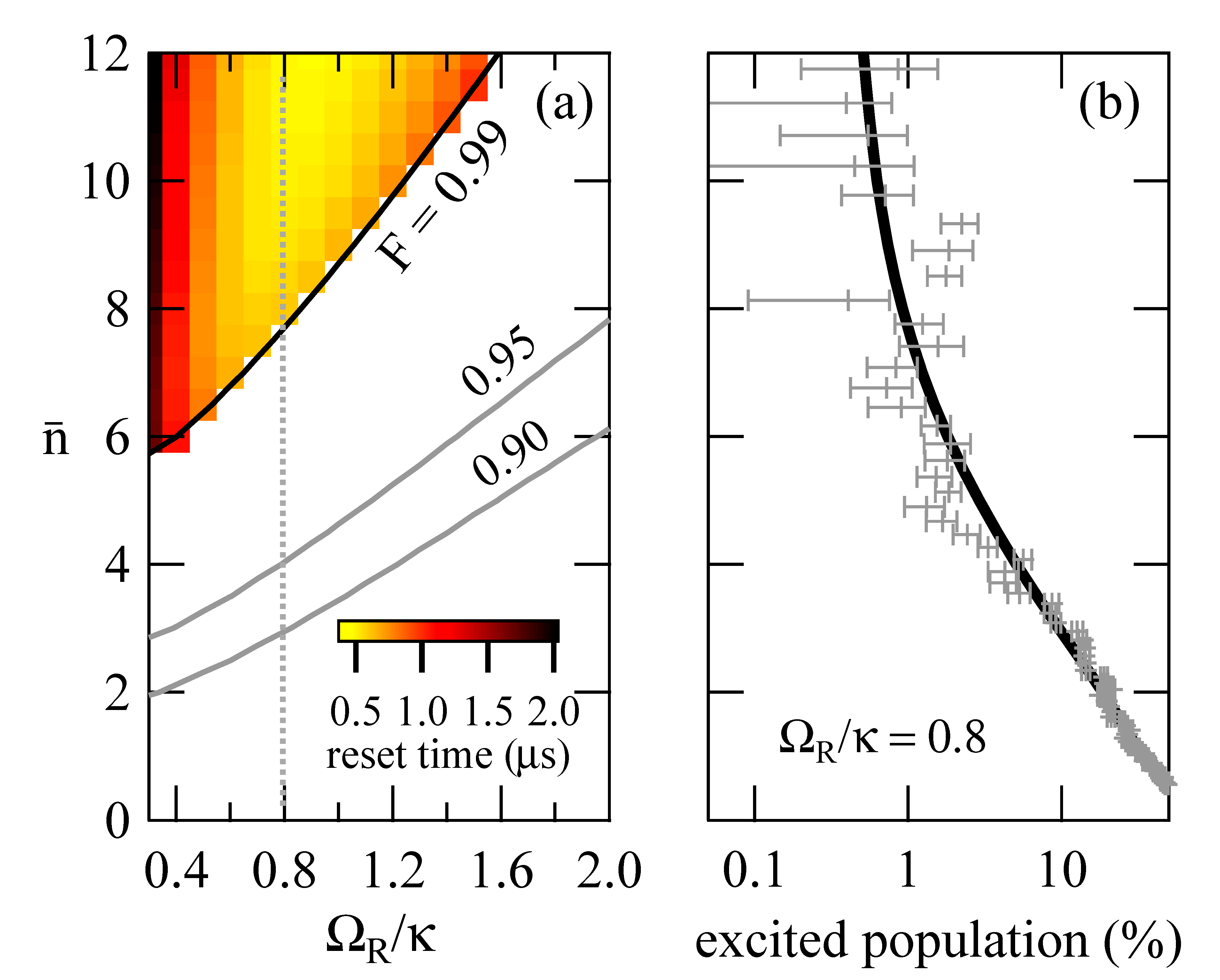}
\caption{(a) Contours of 90, 95, and 99$\%$ predicted ground state preparation fidelity from numerical simulations vs two Rabi drive amplitudes expressed as $\Omega_R/\kappa$ and $\bar{n}$.  For fidelities greater than 99$\%$, the shaded area indicates reset time. (b) Measured excited state population from RPM method (crosses with error bars) compared to numerical simulation (solid line) vs $\bar{n}$ for $\Omega_R/\kappa$ = 0.8.  This population decreases monotonically with $\bar{n}$.}
\label{fig:fidelity_and_time_vs_parameters}
\end{figure}

	In order to optimize the ground state preparation fidelity of DDROP, we performed numerical simulations of the expected fidelity $F$ versus qubit drive amplitude and average cavity excitation, $\Omega_R$ and $\bar{n}$, respectively.  We numerically simulated the Lindblad master equation obeyed by the qubit-cavity density operator, including the two drives and decoherence for both the qubit and cavity, while choosing the initial state to be the cavity in vacuum and an equilibrium state for the qubit.  The dependence of $F$ on $\Omega_R$ for fixed $\bar{n}$ was found to be weak, and fidelities above 99$\%$ were found for $\Omega_R/\kappa >$ 0.3.  Our numerical simulations show that $F$ increases monotonically with $\bar{n}$ for a fixed $\Omega_R$ and that with a higher $\Omega_R$, higher $\bar{n}$ is required to reach the same fidelity, as shown by the contours of constant fidelity in Fig. \ref{fig:fidelity_and_time_vs_parameters}(a).  The simulations did not account for self-Kerr effects that will reduce the fidelity at photon numbers much higher than the range shown.  While 99$\%$ fidelity is reached with a wide range of parameters, reset time is optimized when $\Omega_R \simeq \kappa$, yielding reset times comparable with those of two-qubit gates in the cQED architecture \cite{Paik2012}.  The reset time for the ground state population to reach 99$\%$ is shown by the colored pixels of Fig. \ref{fig:fidelity_and_time_vs_parameters}(a).

	With the guidance provided by these simulations and using RPM to experimentally quantify the fidelity, we have studied DDROP for a wide range of $\Omega_R$ and $\bar{n}$.  The pulse duration was kept fixed at the value 5~$\mu$s, chosen from simulation, to ensure DDROP has reached equilibrium in all conditions.  Fidelities greater than 99$\%$ were achieved for $\Omega_R$ as low as 0.3~$\kappa$ and as high as 1.0~$\kappa$, for $8 \le \bar{n} \le 50$.  For fixed $\Omega_R$ = 0.8~$\kappa$, Fig. \ref{fig:fidelity_and_time_vs_parameters}(b) shows measurement (markers) vs simulation (line) of remaining excited state population vs $\bar{n}$.  Excited state population drops monotonically with $\bar{n}$, in good agreement with numerical simulation.  On the other hand, above approximately $\bar{n}$ = 50 (data not shown), the reset excited state population increased significantly.  This is understood to be due to the breakdown of the dispersive approximation.  Overall, both drive amplitude parameters $\Omega_R$ and $\bar{n}$ have a wide range for which DDROP works well, making it a very reliable and stable protocol.

	As mentioned before, all of the DDROP characterization measurements included a 1~$\mu$s (20~$\kappa^{-1}$) wait between drive pulses and the RPM measurement, to allow the cavity photons to decay.  Therefore, the qubit excited state population begins returning to its equilibrium value as soon as the reset drives are turned off.  This re-equilibration should occur on a timescale given by the mixing time $T_1$, and this is what is found experimentally.

	DDROP is not the first demonstrated qubit reset mechanism to work on superconducting qubits; several distinct methods have been shown previously, including: sideband cooling through higher energy levels \cite{Valenzuela2006}, sweeping the qubit frequency into resonance with a low-Q cavity \cite{Reed2010,Mariantoni2011}, a feedback loop with conditional coherent driving \cite{Riste2012}, and strong projective measurements \cite{Johnson2012,Riste2012b,Campagne2012}.  However, DDROP has many advantages when compared to each of these processes.  First, there is no need to tune in real time the qubit frequency, which means DDROP will still work with fixed-frequency qubits.  There is no need for fast external feedback of any kind, thus simplifying the required setup.  There is also no need for high-fidelity, single-shot readouts or in fact a low-noise amplifier at all.  Finally, the decisive qualitative advantage is that the sensitivity to the drive amplitudes is low, and there is no need for accurate pulse timing or shapes; DDROP can be quickly tuned to near-optimum parameters.

	While a qubit reset is a fundamental primitive necessary for quantum information algorithms, DDROP additionally deals with ``hot" qubits, which are often observed \cite{Corcoles2011}.  While usually unintentional, high qubit temperatures may be beneficial if loss is dominated by dielectrics \cite{Barends2008} or if lower transition frequencies are found to be needed.

	A discussion of the nuance between cooling and reset is now in order.  Qubit reset is ground state preparation with a minimum required fidelity in the shortest possible time, whereas qubit cooling reduces the excited state population below that produced by contact with the external bath.  As shown in this Letter, DDROP satisfies both definitions, yet it differs significantly from other dynamical cooling procedures.  These methods, inherited from their counterpart in atomic physics \cite{Leibfried2003}, have been recently demonstrated in both nanomechanical systems \cite{Chan2011,Riviere2011,Teufel2011} and superconducting qubits \cite{Valenzuela2006,Grajcar2008,Murch2012,Masluk2012}.

	As mentioned earlier, the DDROP protocol is one particular implementation of a wide class of procedures called reservoir engineering or autonomous feedback.  In general, reservoir engineering involves designing the decoherence landscape seen by the qubit with the goal of stabilizing a particular state or manifold.  In the case of DDROP, the stabilized state is $\mid$$g,\alpha\rangle$, whereas in Ref. \cite{Murch2012} the stabilized state is ($\mid$$g\rangle + \mid$$e\rangle$)/$\sqrt[]{2}$, which requires a well-calibrated $\pi/2$ pulse to prepare $\mid$$g\rangle$.  Interestingly, by simply changing the cavity drive frequency to $f_c^g - \chi$, the stabilized state of DDROP becomes $\mid$$e,\alpha\rangle$ instead of $\mid$$g,\alpha\rangle$.  Alternate reservoir engineering schemes can be used to stabilize Bell states of a two-qubit system or to perform an autonomous bit flip quantum error correction in a three-qubit system.  The agreement of DDROP measurements with numerical predictions provides a confirmation of the validity of the basic methods of reservoir engineering and opens the door to many interesting autonomous feedback experiments.

	In conclusion, the DDROP protocol for qubit reset has been experimentally demonstrated on a transmon in a three-dimensional cavity to produce a fast, high-fidelity ground state preparation.  This process satisfies the demand for qubit reset as part of an algorithm, and can also be used to improve the speed and fidelity of ground state preparation over that given by a return to equilibrium.  We have evaluated the performance of the DDROP protocol by using a new method (RPM) for quantifying the excited to ground state population ratio.  The use of DDROP allowed experiments on this qubit to repeat at a rate 60 times faster than waiting 5~$T_1$.  Regardless of initial state, a ground state preparation fidelity of 99.5$\%$ was achieved in less than 3~$\mu$s.  Simulation predicts higher fidelities are possible; for example, simply reducing $P_e$ from 9$\%$ to 1$\%$ and using $\bar{n} = 25$, simulations predict a fidelity of 99.99$\%$.  The requirements and constraints of DDROP are fewer than other forms of reset; neither feedback, high-fidelity readout nor qubit tunability are necessary.  DDROP is readily applicable and practically useful for most cQED systems.

	This research was supported by IARPA under Grant No. W911NF-09-1-0369, ARO under Grant No. W911NF-09-1-0514 and NSF under grants Grant No. DMR-1006060 and No. DMR-0653377.  Z.L. and M.M. acknowledge partial support from French Agence Nationale de la Recherche under the project EPOQ2, No. ANR-09-JCJC-0070.  Facilities used were supported by the Yale Institute for Nanoscience and Quantum Engineering and NSF Grant No. MRSEC DMR 1119826.

\bibliographystyle{apsrev}
%\bibliography{resetPaperBib}

\end{document}